\documentclass[conference]{IEEEtran}
\IEEEoverridecommandlockouts

\def\BibTeX{{\rm B\kern-.05em{\sc i\kern-.025em b}\kern-.08em
    T\kern-.1667em\lower.7ex\hbox{E}\kern-.125emX}}
\ifCLASSINFOpdf
\fi
\usepackage{verbatim}
\usepackage{algorithm}
\usepackage{graphicx}
\usepackage{cite}
\usepackage{amssymb,amsfonts}
\usepackage{algorithmicx}
\usepackage[noend]{algpseudocode}
\usepackage{setspace}
\usepackage{float}
\usepackage{subfigure}
\usepackage{epstopdf}
\usepackage{amsthm}
\newtheorem{lemma}{Lemma}
\newtheorem{theorem}{Theorem}
\usepackage{stfloats}
\usepackage{enumerate}
\usepackage{multirow}
\usepackage{latexsym}
\usepackage{multicol,lipsum}
\usepackage{cuted} 

\usepackage{tikz}
\usepackage{textcomp}
\usepackage{xcolor}

\pdfoutput=1

\usepackage[cmex10]{amsmath}
\usepackage[cmex10]{mathtools}

\algdef{SE}[DOWHILE]{Do}{doWhile}{\algorithmicdo}[1]{\algorithmicwhile\ #1}%
\allowdisplaybreaks
\makeatletter 
\def\@eqnnum{{\normalsize \normalcolor (\theequation)}} 

\makeatother 
\begin{document}
\title{Securing V2I Backscattering from Eavesdropper
\vspace{-3pt}}
\author{\IEEEauthorblockN{Ruotong Zhao, Deepak Mishra, and Aruna Seneviratne}
\IEEEauthorblockA{School of Electrical Engineering and Telecommunications, University of New South Wales, Sydney, NSW 2052, Australia\\Emails:   ruotong.zhao@student.unsw.edu.au,\,d.mishra@unsw.edu.au,\,a.seneviratne@unsw.edu.au
\vspace{-5mm}
}}
\maketitle

\begin{abstract}
As our cities become more intelligent and more connected with new technologies like 6G, improving communication between vehicles and infrastructure is essential while reducing energy consumption. This study proposes a secure framework for vehicle-to-infrastructure (V2I) backscattering near an eavesdropping vehicle to maximize the sum secrecy rate of V2I backscatter communication over multiple coherence slots. This sustainable framework aims to jointly optimize the reflection coefficients at the backscattering vehicle, carrier emitter power, and artificial noise at the infrastructure, along with the target vehicle's linear trajectory in the presence of an eavesdropping vehicle in the parallel lane. To achieve this optimization, we separated the problem into three parts: backscattering coefficient, power allocation, and trajectory design problems. We respectively adopted parallel computing, fractional programming, and finding all the candidates for the global optimal solution to obtain the global optimal solution for these three problems. Our simulations verified the fast convergence of our alternating optimization algorithm and showed that our proposed secure V2I backscattering outperforms the existing benchmark by over $4.7$ times in terms of secrecy rate for $50$ slots. Overall, this fundamental research on V2I backscattering provided insights to improve vehicular communication's connectivity, efficiency, and security.
\end{abstract}


\section{Introduction}
The cutting-edge advancements in 6G technology have opened up exciting possibilities for smart cities that enhance communication between vehicles and infrastructure - a crucial aspect of autonomous driving systems~\cite{V2I}. However, as the Internet of Things (IoT) device count continues to soar, adopting a sustainable approach that reduces energy consumption in vehicular communications is imperative~\cite{GC23}. This is where backscatter technology comes in, a low-power communication paradigm that has captured the attention of experts and is poised to transform the IoT landscape~\cite{gu2023act}. To ensure seamless V2I communication, we must address security concerns like eavesdropping - where malicious actors attempt to intercept sensitive signals~\cite{mensi2021pls}. Thus, backscatter systems have emerged as a sustainable solution, modulating continuous wave (CW) signals to enable energy-efficient V2I communication that will shape the future of autonomous driving.
These backscattering tags are integral to vehicular systems, facilitating reliable signal backscattering~\cite{khan2021backscatter}, yet they face challenges due to the limited range of backscatter communication. 
Recent strategies in resource allocation show potential in overcoming these limitations and boosting network performance~\cite{khan2021learning}. 

\subsection{State-of-the-Art}
Research on improving security in backscatter-assisted vehicular communications has given more importance to physical layer security (PLS) than key encryption methods because the former is better suited to the dynamic and resource-constrained vehicular networks~\cite{zhao2022securing}. The authors in~\cite{makarfi2020physical} conducted a study on the PLS of a vehicular network that uses a reconfigurable intelligent surface to investigate the average capacity. The work in~\cite{saad2014physical} suggests that the reader can intentionally transmit artificial noise (AN) with CW signals to protect against eavesdropping in the reflection link. This is possible because the reader can partially eliminate it by using successive interference cancellation (SIC) since it is the emitter and knows the transmitted signal by itself. But the eavesdropper cannot distinguish it. This concept has been extended by Yang et al.~\cite{yang2020exploiting}, who use receiver-side random noise, and by Li et al.~\cite{li2022improving}, who propose source AN injection to secure backscatter-assisted vehicular-to-pedestrian communication links. In addition to resource allocation, the vehicle trajectory is also identified as a crucial factor in vehicular communications. It influences security and system performance~\cite{wang2019backscatter}.

\subsection{Motivation and Contributions}
To the best of our knowledge, no investigation has been conducted on protecting V2I backscattering from eavesdropping attacks despite the advancements in the literature on backscatter-enabled secure vehicular networks. This paper aims to bridge this gap by presenting a comprehensive solution for V2I communications that can help achieve safe sustainable transportation. The proposed solution has the potential to secure low-cost, sustainable sensing and communication applications in V2I networks for intelligent transportation systems. These goals of realizing secure green V2I backscatter communications are the following key contributions:
\begin{itemize}
\item A new V2I backscattering protocol is proposed to secure communication between a tag-equipped vehicle and a monostatic reader-equipped infrastructure. This green, low-cost, and sustainable protocol protects V2I backscattering from eavesdropping attacks by adjacent vehicles.
\item We optimize the infrastructure's transmit power, tag's reflection coefficient, and vehicle's linear trajectory to realize the maximum secrecy rate for V2I backscattering.
\item As the principal problem is non-convex, we utilize alternating optimization (AO) to break it into three subproblems, which can then be solved globally optimally. Next, we adopt parallel computing for reflection coefficient, fractional programming (FP) for power control, and identify all possible candidates for the global optimal solution in the trajectory design problem. This method provides a low-complexity approach to obtain a near-global optimal solution for the joint optimization problem.
\item Our theoretical propositions are validated through extensive numerical simulations that provide nontrivial design insights for different system parameters and demonstrate superior performance compared to benchmark methods.
\end{itemize}

\section{System Description}\label{sec:System}
\subsection{V2I setup and Channel Model}\label{sec:network}
\begin{figure}[htbp]
\centerline{\includegraphics[width=3.4in]{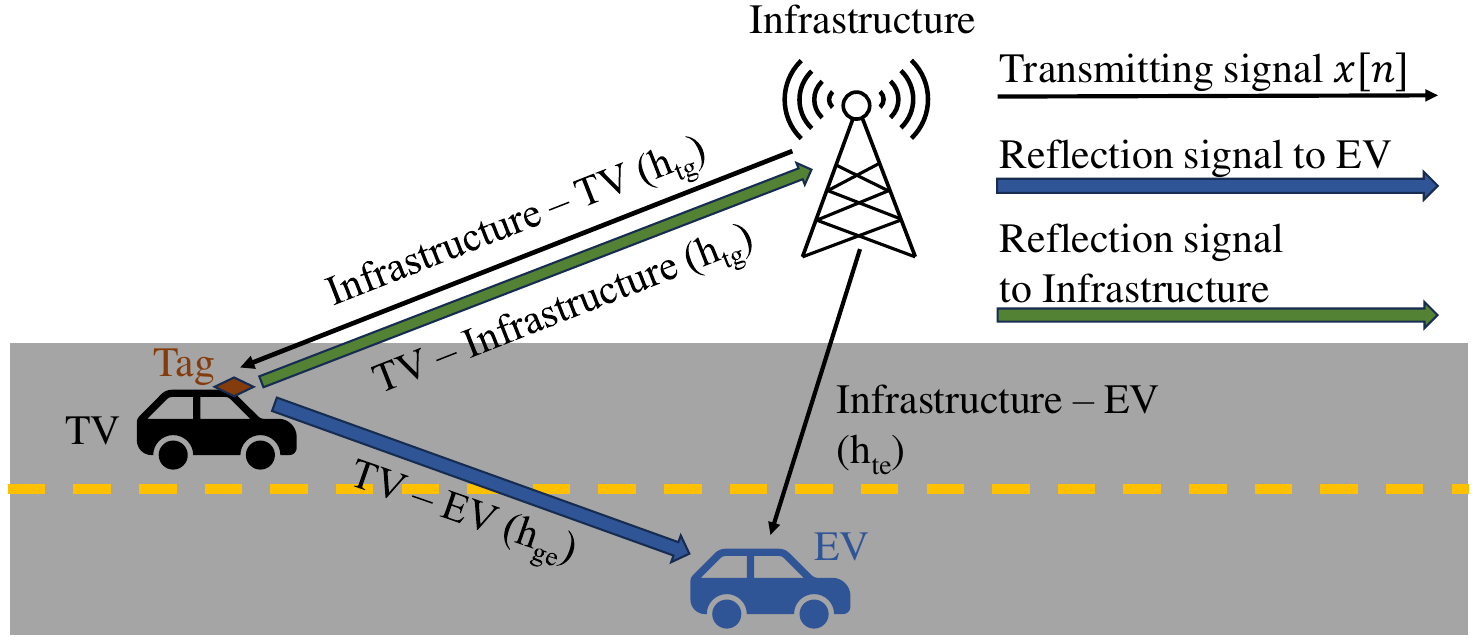}}
\caption{Considered backscattering-aided V2I system model showing $2$ vehicles situated in $2$ distinct lanes.}
\label{fig:model}
\end{figure}

As shown in Fig.~\ref{fig:model}, we investigate a system where a tag-equipped target vehicle (TV) and an eavesdropper vehicle (EV) are situated in two distinct lanes. 
We consider quasi-static large-scale fading in this work, where the entire time is divided into \(N\) small time slots \(\mathcal{N} \in \{1,2, ..., N\}\), and the duration of each time slot is $t$. The TV is expected to cover a minimum distance of \(L\) m, and the EV is assumed to maintain a consistent speed \(v_c\), where the positions of the EV and infrastructure are known. The position of the infrastructure access point, acting as a mono-static reader \cite{9048853}, is denoted by  $3$ dimensional coordinates \(t = (x_t, y_t, h)\). Meanwhile, the EV's and TV's positions are \(e[n] = (x_e[n], y_e)\) and \(g[n] = (x_g[n], y_g)\), respectively. Therefore, the position of the tag we optimize here is a one-dimensional variable, \textbf{$x_g$}.  

In this analysis, the instantaneous velocity and acceleration of the TV in the $n$th slots are constrained by the interval $[v_{\min}, v_{\max}]$, and $[a_{\min}, a_{\max}]$ respectively. The distance between the infrastructure and the EV during the \( n \)th slot is \( d_{te}[n], n \in \mathcal{N} \). Additionally, the separation of infrastructure-TV during the \( n \)th time slot can be expressed as:
\begin{equation}\label{dis_tg}
\begin{aligned} 
    d_{tg} \hspace{-0.4mm}[n] \hspace{-1.2mm}=\hspace{-1.5mm} \sqrt{\hspace{-0.8mm}(\hspace{-0.6mm}x_t\hspace{-1mm} - \hspace{-1mm}x_g\hspace{-0.5mm}[n]\hspace{-0.3mm})\hspace{-0.2mm}^2 \hspace{-0.8mm}+ \hspace{-0.8mm}(\hspace{-0.5mm}y_t \hspace{-1mm}- \hspace{-0.9mm}y_g\hspace{-0.5mm})^2\hspace{-1mm} +\hspace{-1mm}h\hspace{-0.2mm}^2}\hspace{-1mm}= \hspace{-1.3mm}\sqrt{\hspace{-0.8mm}x_g\hspace{-0.5mm}[n] ^2 \hspace{-1mm}- \hspace{-1mm}2 x_t x_g\hspace{-0.5mm}[n] \hspace{-1mm}+\hspace{-1mm} K_{tg}}.
    \end{aligned}
\end{equation}
Similarly, the separation of TV-EV in the \(n\)th time slot is:
\begin{equation}\label{dis_ge}
\begin{aligned}
    d_{ge}\hspace{-0.5mm}[n] \hspace{-1.3mm}=\hspace{-1.6mm} \sqrt{\hspace{-0.9mm}(\hspace{-0.7mm}x_e\hspace{-0.5mm}[n]\hspace{-1.1mm} - \hspace{-1mm}x_g\hspace{-0.6mm}[n]\hspace{-0.5mm})\hspace{-0.3mm}^2 \hspace{-1.05mm} + \hspace{-0.95mm}(\hspace{-0.55mm}y_e \hspace{-1.3mm}- \hspace{-0.95mm}y_g\hspace{-0.7mm})\hspace{-0.35mm}^2\hspace{-0.8mm}}\hspace{-0.5mm}= \hspace{-1.5mm}\sqrt{\hspace{-0.95mm}x_g\hspace{-0.35mm}[n] \hspace{-0.5mm}^2 \hspace{-1.2mm}- \hspace{-1.2mm}2 x_e \hspace{-0.5mm} [n] \hspace{-0.4mm}x_g\hspace{-0.5mm}[n] \hspace{-1.1mm}+\hspace{-1.1mm} K_{ge}\hspace{-0.5mm}[n]},
\end{aligned}
\end{equation}
where $K_{tg} \hspace{-1mm} = (y_t - y_g)^2 + h^2 + x_t^2$, $K_{ge}[n] = (y_e - y_g)^2 + x_e [n]^2$.
Moreover, we assume that the infrastructure, TV, and EV are equipped with a single antenna. The channel gain coefficients of TV-EV,  infrastructure-EV, and infrastructure-TV are represented by \( h_{ge}[n] \sim \mathbb{CN} \left(0, \frac{s_{ge}[n]}{d_{ge}^i[n]}\right) \),  \( h_{te}[n] \sim \mathbb{CN}\left(0,\frac{s_{te}[n]}{d_{te}^i[n]}\right) \), and \( h_{tg}[n] \sim \mathbb{CN}\left(0, \frac{s_{tg}[n]}{d_{tg}^i[n]}\right) \) respectively. Here, \( i \) signifies the path-loss exponent, and $s_{te}$ is the distance-independent path loss constant of infrastructure-EV link. 
The additive white Gaussian noise (AWGN) at the EV and infrastructure are represented as \( m_e [n] \sim \mathbb{CN}(0, \sigma^2_e) \) and \( m_t [n] \sim \mathbb{CN}(0, \sigma^2_t) \), respectively.
We assume the channel state information (CSI) for both \( h_{te} \) and \( h_{tg} \) is available at the infrastructure through channel reciprocity according to the literature~\cite{CSI2019}\cite{ming2023improving}. 

\subsection{Transmission Signal Analysis}\label{sec:SignalAnal}
We consider $x [n]\in \mathbb{C} \  \text{and} \ c[n] \in \mathbb{C}$ to be the CW of the infrastructure and the information signal of tag at the $n$th slot, respectively, where $\left|c[n]\right|^2 = \left|x[n]\right|^2 = 1$. Moreover, $x [n] = s[n] + z[n], \forall n \in 
\mathcal{N}$, with $s [n] \in \mathbb{C}$ being the CW signal, and $z[n] \in \mathbb{C}$ representing the AN. 
In this system, $\mathbf{p}_s$ and $\mathbf{p}_a$ represent the vectors of transmit powers across $N$ slots for the CW and AN, $\mathbf{p}_s = \left\{ p_s[1], \ldots, p_s[N] \right\}$ and $\mathbf{p}_a = \left\{ p_a[1], \ldots, p_a[N] \right\}$. Here $p_s[n] = \mathbb{E}\{s[n]^2\}$ and $p_a[n] = \mathbb{E}\{z[n]^2\}$ correspond to the powers allocated to the CW and AN in the $n$th slot, where $\mathbb{E}\{\cdot\}$ signifies the expectation operator.
The total power budget is denoted as $P$, then the power constraint can be articulated as:
\begin{equation}
    \sum_{n = 1}^{N}p_s [n] + \sum_{n = 1}^{N}p_a [n] \leq P
\end{equation}
Consequently, the received signal at the infrastructure with perfect CSI by considering the transmitting powers is:
\begin{equation}\label{reci_main}
\begin{aligned}
    y_t [n] &= h_{tg} [n]^2 \sqrt{p_s [n]\beta [n]} c[n] + m_t[n]\\ &+ \sqrt{\alpha} h_{tg} [n]^2 \sqrt{p_a[n]\beta [n]}c[n],
\end{aligned}
\end{equation}
where $\sqrt{\beta [n]}$ denotes the reflection coefficient at $n$th time slot and $\alpha \in [0,1]$ is the
attenuation factor which denotes how successful the reader is
in cancelling the backscattered AN~\cite{MIMO_backscat}. Note that unlike the infrastructure, the EV cannot perform AN attenuation due to the absence of prior knowledge. As the eavesdropper receives the superposition of the CW signal from the reader and the backscattered signal from the tag, we assume that the CW signal is known to the eavesdropper and thus can be easily removed from the EV's received signal. Thus, the eavesdropper's received signal is:
\begin{equation}\label{reciv_e}
\begin{aligned}
    y_e [n] &= h_{ge} [n]  h_{tg}[n]  \sqrt{p_s[n]\beta [n]} c[n] + m_e[n] \\
    & +h_{ge} [n] h_{tg}[n]  \sqrt{p_a[n]\beta [n]}c[n] + \sqrt{p_a[n]}h_{te}[n].
    \end{aligned}
\end{equation}

\section{Problem Definition}\label{sec:Problem}
\subsection{Secrecy Rate Expressions}
Before drawing the secrecy rate for this proposed V2I backscattering model, we will first analyse the Signal-to-Interference-plus-Noise Ratio (SINR) for infrastructure-TV and TV-EV links.  
By considering the recived signals in~\eqref{reci_main} and~\eqref{reciv_e} , the SINR of the infrastructure-TV link at the $n$th time slot can be expressed as:
\begin{equation}\label{SINR_main}
\begin{aligned}
    \gamma_t [n] \hspace{-1mm}
    &=\hspace{-1mm}\frac{p_s [n]\mathrm{E}\{|h_{tg}[n]|^2\}^2 \beta [n]}{\alpha p_a [n]\mathrm{E}\{|h_{tg}[n]|^2\}^2 \beta [n] + \sigma_r^2}\\
    &= \hspace{-1mm}\frac{p_s [n]|s_{tg}[n]|^4 \beta [n] (x_g[n] ^2 - 2 x_t x_g[n] + K_{tg})^{-i}}{\alpha \beta [n] p_a[n]|s_{tg}[n]|^4 (x_g[n] ^2 \hspace{-0.8mm} - \hspace{-0.8mm} 2 x_t x_g[n]\hspace{-0.8mm} + \hspace{-0.8mm} K_{tg})^{-i} \hspace{-0.8mm} + \hspace{-0.8mm} \sigma_r^2},
\end{aligned}
\end{equation}
And the SINR of the TV-EV link at the $n$th time slot is:
\begin{equation}\label{SINR_eave}
\begin{aligned}
    \gamma_e [n] \hspace{-0.8mm}& = \frac{\mathrm{E}\{|h_{ge} [n]|^2\}  \mathrm{E}\{|h_{tg}[n]|^2\}  p_s[n]\beta [n]}{\frac{\mathrm{E}\{|h_{ge} [n]|^2\} p_a[n]\beta[n]}{\mathrm{E}\{|h_{tg}[n]|^2\}^{-1}} + p_a[n]\mathrm{E}\{|h_{te}[n]|^2\}+\sigma_e^2} \\&=\hspace{-0.8mm}\frac{|s_{ge} [n]|^2 |s_{tg}[n]|^2\beta [n] p_s[n] D [n]^{-1} }{\frac{|s_{ge} [n]|^2 |s_{tg} [n]|^2 \beta [n]p_a[n]} {D[n]}+|s_{te} [n]|^2 p_a[n]d_{te} [n]^{-i} + \sigma_e^2},
\end{aligned}
\end{equation}
where $D [n] = (x_g[n] ^2 - 2 x_e [n] x_g[n] + K_{ge}[n])^{i/2} (x_g[n] ^2 + 2 x_s x_g[n] + K_{tg})^{i/2}, \forall n \in \mathcal{N}$.

Noting the SINR definitions from~\eqref{SINR_main} and~\eqref{SINR_eave}, the rates or spectral efficiencies in bits/second/Hz ((bps/Hz))~\cite{zhao2022securing} for the legal backscattering link $R_{t}$ and eavesdroppering link $R_{e}$ for the $n$th time slot can be written as:
\begin{equation}
    R_{t} [n] = \log_2 \left(1+\gamma_{t}[n]\right),\quad
    R_{e} [n] =\log_2 \left(1+\gamma_{e}[n] \right).
\end{equation}
Using the Wyner wiretap channel model~\cite{zhao2022securing} along with the above expressions, the secure rate of the $n$th time slot is:
\begin{equation}\label{R} 
\begin{aligned}
R [n] \hspace{-0.7mm}= \hspace{-0.7mm}\left\{\hspace{-0.6mm}R_{t} [n]\hspace{-0.6mm}-\hspace{-0.6mm}R_{e}[n]\right\}^+\hspace{-1mm}=\hspace{-0.6mm}{\max}\hspace{-0.8mm}\left\{\hspace{-0.8mm}\log_{2}\hspace{-0.6mm}\left(\hspace{-0.6mm}\frac{1+\gamma_{t}[n]}{1+\gamma_{e}[n]}\hspace{-0.6mm}\right), 0\hspace{-0.8mm}\right\}.
\end{aligned}
\end{equation}
Then the proposed optimization problem is formulated as: \begin{align}
\mathcal{P}_1:& \underset{\mathbf{p}_a, \mathbf{p}_s, s_g [n],\beta[n], \forall n \in \mathcal{N}}{\text{maximize}}  \sum_{n=1}^{N} R [n] \quad \text{subject to (s.t.) :} \label{formulation} \nonumber \\
C1 :& \ \beta [n] \in (0,1),  \forall n \in \mathcal{N}, \nonumber \\
C2 :& \ (1 - \beta[n])|h_{tg}[n]|^2 \geq E_b, \forall n  \in \mathcal{N}, \nonumber \\
C3 :& \ \sum_{n = 1}^{N}p_s [n] + \sum_{n = 1}^{N}p_a [n] \leq P, \nonumber\\
C4 :& \ x_g [0] = 0, \qquad C5 : \ x_g [N] \geq L, \nonumber \\
C6 :& \ v_\text{min} \leq v_t[n] \leq v_\text{max}, \forall n \in \mathcal{N}, \nonumber \\
C7 :& \ a_{\min} \hspace{-1mm} \leq \hspace{-1mm} \frac{2\hspace{-0.7mm}\left(\hspace{-0.5mm}x_g[n] \hspace{-0.8mm} - \hspace{-0.8mm} x_g [n-1] \hspace{-0.8mm} - \hspace{-0.8mm} v_t [n-1]t\hspace{-0.3mm}\right)}{t^2} \hspace{-1mm} \leq \hspace{-1mm}a_{\max}, \forall n \in \mathcal{N}.\nonumber
\end{align}
where the reflection coefficient \(\sqrt{\beta [n]}\) is a real value lying \([0,1]\). Constraint~({\rm C2}) ensures the tag's minimum power threshold, and ({\rm C3}) governs the power budget. The initial location of the TV is \(0\), and it should cover a minimum distance of \(L\), mandated by~({\rm C4})-({\rm C5}). Furthermore, the system imposes the velocity and acceleration range in the last $2$ constraints.

\subsection{Proposed 
Methodology}\label{sec:methdology}
\begin{figure}[htbp]
\centerline{\includegraphics[width=2.6in]{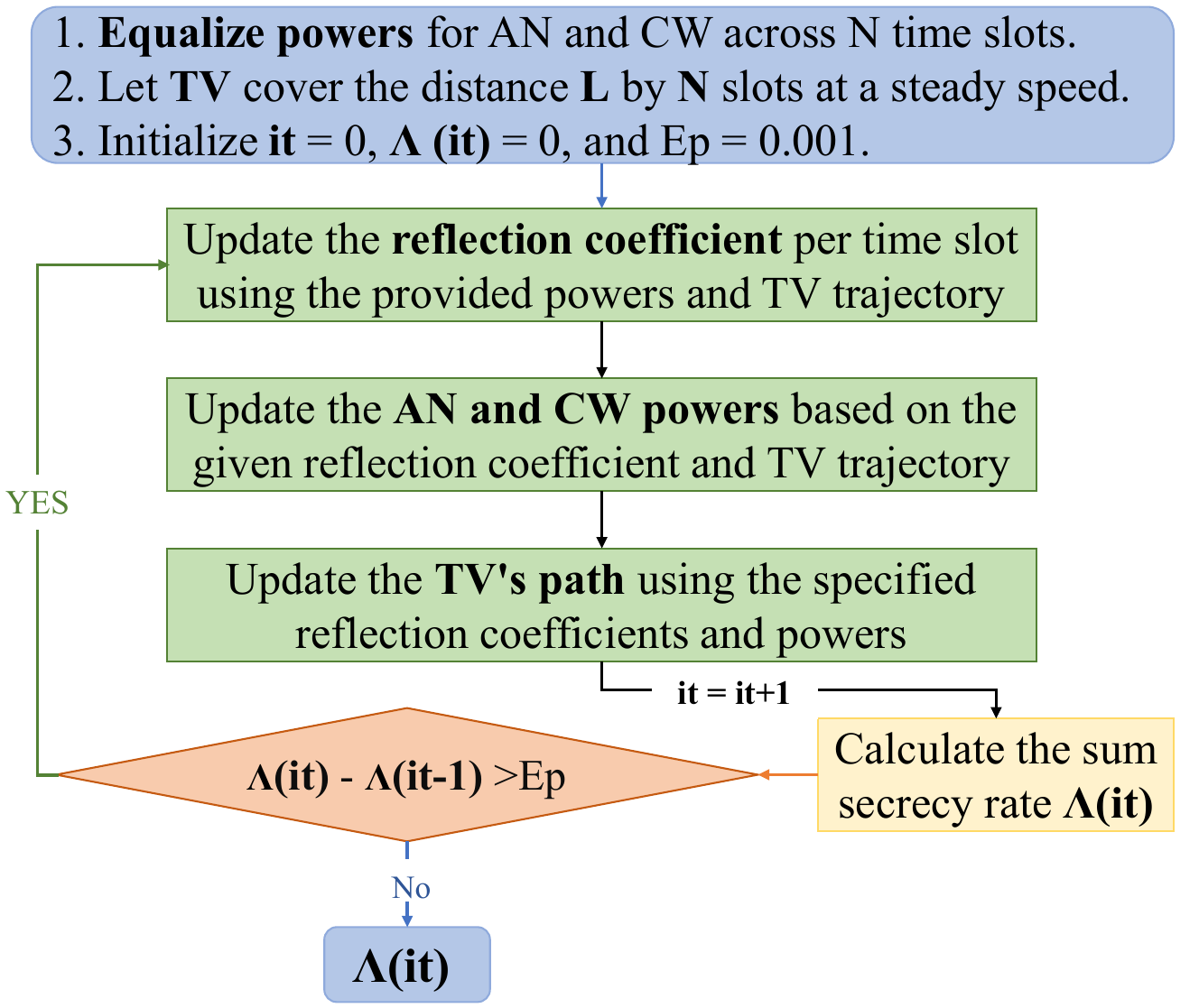}}
\vspace{-3pt}
\caption{The flowchart of the proposed methodology}
\label{fig:flowchart}
\vspace{-5pt}
\end{figure}

We employ the AO method to decouple the principle problem into reflection coefficient optimization, CW and AN power control, and vehicle trajectory design. Initially, the reflection coefficient is optimized for each time slot based on equal CW and AN transmission powers, \(p_a[n] = p_s[n] = \frac{P}{2N}\), and a uniform speed for TV, \(v_t[n] = \frac{L}{N}\), across all \(N\) slots. 
Following this, power allocation is adjusted, accounting for the given reflection coefficients and TV trajectory. Then, vehicle position optimization proceeds, dependent on the optimal reflection coefficients and power levels. 
This iterative loop repeats will repeat from the reflection coefficient optimization until convergence is achieved to obtain optimal results. The algorithmic specifics are detailed in Fig.~\ref{fig:flowchart}.


\section{PROPOSED SECRECY RATE OPTIMIZATION}\label{sec:Solution}
\subsection{Backscattering Coefficient Optimization}\label{sec:bate_opt}
\begin{figure*}[!t]
\setcounter{equation}{10}
\begin{equation}\label{trans}
\begin{aligned}
R^* [n] = &\log_2\left(2 y_1[n] \sqrt{p_s [n]|s_{tg}[n]|^4 \beta [n] (x_g[n] ^2 - 2 x_t x_g[n] + K_{tg})^{-i} + \alpha \beta [n] p_a[n]|s_{tg}[n]|^4 (x_g[n] ^2 \hspace{-0.8mm} - \hspace{-0.8mm} 2 x_t x_g[n]\hspace{-0.8mm} + \hspace{-0.8mm} K_{tg})^{-i} \hspace{-0.8mm} + \hspace{-0.8mm} \sigma_r^2} \right.\\
& \left.- y_1 [n]^2 \left(\frac{|s_{ge} [n]|^2 |s_{tg}[n]|^2\beta [n] p_s[n]} {D [n]}+\frac{|s_{ge} [n]|^2 |s_{tg} [n]|^2 \beta [n]p_a[n]} {D[n]}+|s_{te} [n]|^2 p_a[n]d_{te} [n]^{-i} + \sigma_e^2\right)\right) \\ &+\log_2\hspace{-1mm} \left(\hspace{-1mm}2 y_2[n]\sqrt{\frac{|s_{ge} [n]|^2 |s_{tg} [n]|^2 \beta [n]p_a[n]} {D[n]}+|s_{te} [n]|^2 p_a[n]d_{te} [n]^{-i} + \sigma_e^2} - y_2 [n]^2\hspace{-1mm} \left(\hspace{-1mm}\frac{\alpha \beta [n] p_a[n]|s_{tg}[n]|^4}{(x_g[n] ^2 \hspace{-0.8mm} - \hspace{-0.8mm} 2 x_t x_g[n]\hspace{-0.8mm} + \hspace{-0.8mm} K_{tg})^{i}} \hspace{-0.8mm}+ \hspace{-0.8mm}\sigma_r^2\hspace{-1mm}\right)\hspace{-1mm}\right).
\end{aligned}
\end{equation}
\noindent\rule{18cm}{0.5pt}
\normalsize
\end{figure*}
As the previous discussion in Section~\ref{sec:methdology}, we will first adopt parallel computing to optimal the reflection coefficient with the given transmit powers of infrastructure and trajectory of TV. Recall the problem in $\mathcal{P}_{1}$ and noting the reflection coefficient optimization problem is non-convex, here we adopt the FP transformation in the literature~\cite{FPMethods} for the secrecy rate $R[n]$ to $R^*[n], \forall n \in \mathcal{N}$ in~\eqref{trans}, where the $y_1 [n]$ and $y_2 [n]$ are the intermediary variables and can be expressed as: 
\begin{equation}\label{y1}
\begin{aligned}
    &y_1 [n] \hspace{-1mm}= \hspace{-1.3mm} \frac{\frac{p_s [n]|s_{tg}[n]|^4 \beta [n]} {(x_g[n] ^2 - 2 x_t x_g[n] + K_{tg})^{i}}+ \frac{\alpha \beta [n] p_a[n]|s_{tg}[n]|^4} {(x_g[n]^2 - 2 x_t x_g[n]+ K_{tg})^{i}}+  \sigma_r^2}{\frac{|s_{tg}[n]|^2\beta [n] p_s[n]} {D [n]|s_{ge} [n]|^{-2}}\hspace{-0.5mm}+\hspace{-0.5mm}\frac{ |s_{tg} [n]|^2 \beta [n]p_a[n]} {D[n]|s_{ge} [n]|^{-2}} \hspace{-0.7mm}+ \hspace{-0.7mm}\frac{|s_{te} [n]|^2 p_a[n]}{d_{te} [n]^{i}} \hspace{-0.7mm} + \hspace{-0.7mm}\sigma_e^2}, 
    \end{aligned}
\end{equation}
\begin{equation}\label{y2}
\begin{aligned}
    y_2 [n] \hspace{-1mm} = \hspace{-1.3mm} \frac{\frac{|s_{ge} [n]|^2 |s_{tg} [n]|^2 \beta [n]p_a[n]} {D[n]}+|s_{te} [n]|^2 p_a[n]d_{te} [n]^{-i} + \sigma_e^2}{\frac{\alpha \beta [n] p_a[n]|s_{tg}[n]|^4}{(x_g[n] ^2 - 2 x_t x_g[n] + K_{tg})^{i}} + \sigma_r^2}
    \end{aligned}
\end{equation}
By considering that the reflection coefficients are independent for each time slot, the reflection coefficient optimal problem is formulated as follows:
\begin{align*}
&\mathcal{P}_2: \underset{\beta[n]}{\text{maximize}} \  R^* [n], \
\text{s.t.}:\ ({\rm C1}) \ \text{and} \ ({\rm C2}).
\end{align*}

\begin{lemma}\label{lemma: reflection_conv}
The reflection coefficient optimization problem $\mathcal{P}_{2}$ is a convex problem.
\end{lemma}
\vspace{-3mm}
\begin{proof}
It can be observed that the transformed objective function in~\eqref{trans} comprises the logarithm of the difference between a concave and a linear function. 
It implies the objective in~$\mathcal{P}_2$ is concave as the constraints are linear~\cite{2004convex}.
\end{proof}

According to the Lemma~\ref{lemma: reflection_conv}, the problem $\mathcal{P}_2$ enabled the determination of the global optimum of the reflection coefficients for each discrete time interval,  as delineated in Algorithm~\ref{Algo:bate}.  
The two most pivotal steps involve driving the closed-form expressions for the intermediary variables $y_1[n]$ and $y_2[n], n \in \mathcal{N}$, and solving the convex problem $\mathcal{P}_2$, as clarified in step $5$ and $6$ of Algorithm~\ref{Algo:bate}, respectively. Given that the objective function of $\mathcal{P}_2$ exhibits unimodality, it is amenable to efficient computation using \texttt{Golden Section} method with the lower and upper limit bound by the constraints and the tolerance $\xi$ can be manually selected. It is particularly adept at locating the extremum of unimodal functions within a prescribed interval.
\begin{algorithm}[!t]
\setstretch{1.38}
	{\small\caption{Reflection coefficient optimization}\label{Algo:bate}
		\begin{algorithmic}[1]
			\Require 
			\parbox[t]{\dimexpr\linewidth-\algorithmicindent-\algorithmicindent\relax}{$\alpha, p_s[n], p_a[n], s_{tg}[n], x_g[n], d_{te}[n], D[n], \forall n \in \mathcal{N}$, $L$, $\sigma_r^2, \sigma_e^2$, $x_t$, $K_{tg}$, $i$, and tolerance $\xi$.\strut}
			\Ensure 
			\parbox[t]{\dimexpr\linewidth-\algorithmicindent-\algorithmicindent\relax} {Optimal reflection coefficient ${\beta [n]^*},\forall n \in \mathcal{N}$.\strut}
			\State Set $\mathrm{it}=1,\, \beta [n]^{\rm(it)}=0.5, \forall n \in \mathcal{N}$.
			\State  Calculate $\Lambda^{(\mathrm{1})} = \sum_{n = 1}^N R [n]$ by substituting $\beta [n]^{\rm(it)}, \forall n \in \mathcal{N}$. 
			\Do\Comment{Iteration}
                \vspace{-1mm}
			\State Set $y_1\hspace{-0.7mm}[\hspace{-0.3mm}n\hspace{-0.3mm}]$ and $y_2 \hspace{-0.4mm}[\hspace{-0.3mm}n\hspace{-0.3mm}]$ by substituting $\beta \hspace{-0.3mm}[\hspace{-0.3mm}n\hspace{-0.3mm}]\hspace{-0.4mm}^{(\hspace{-0.3mm}\mathrm{it}\hspace{-0.3mm})}$ into~\eqref{y1}-\eqref{y2}.
            
            \State\parbox[t]{\dimexpr\linewidth-\algorithmicindent-\algorithmicindent\relax}{Set $\mathrm{it}=\mathrm{it}+1$. \strut}
            \State\parbox[t]{\dimexpr\linewidth-\algorithmicindent- 
          \algorithmicindent\relax} {Solve $\mathcal{P}_{2}$ using \texttt{Golden Section} and return $\beta[n] ^{\rm(it)}.$\strut}
			\State 
			\parbox[t]{\dimexpr\linewidth-\algorithmicindent-\algorithmicindent\relax}{Calculate $\Lambda\hspace{-0.5mm}^{(\mathrm{it})}\hspace{-1.8mm} =\hspace{-1.8mm} \sum_{n = 1}^N \hspace{-1mm}R\hspace{-0.3mm} [\hspace{-0.3mm}n\hspace{-0.3mm}]$ by substituting $\beta \hspace{-0.3mm} [\hspace{-0.3mm}n\hspace{-0.3mm}]\hspace{-0.3mm}^{\rm(it)}\hspace{-0.9mm}, \hspace{-0.3mm}\forall\hspace{-0.3mm} n \hspace{-0.7mm}\in \hspace{-0.7mm}\mathcal{N}$.\strut} 
			\doWhile{$\left(\Lambda^{(\mathrm{it})}-\Lambda^{(\mathrm{it}-1)}\right)\le\xi$}\Comment{Termination}
			\State \textbf{Return}  ${\beta[n]^{*}}={\beta[n]^{\rm(it)}, \forall n \in \mathcal{N}}$. 
				
		\end{algorithmic}
	}
\end{algorithm}

\subsection{Optimal Power Allocation}\label{sec:Power_opt}
Upon determination of the optimal reflection coefficients for each time slot, attention is shifted towards optimizing the power allocation. The FP transformation delineated in~\eqref{trans} is employed once more to convert this non-convex power control optimization problem and subsequently expressed as:
\begin{align*}
\mathcal{P}_3: \underset{\textbf{p}_a,\textbf{p}_s}{\text{maximize}} \hspace{-0.5mm} \sum_{n=1}^{N} R^* [n], \quad 
\text{s. t. } ({\rm C3}).
\end{align*}
\begin{lemma}
The power control optimization problem $\mathcal{P}_{3}$ is a convex problem.
\end{lemma}
\vspace{-3mm}
\begin{proof}
The convexity proof for $\mathcal{P}_2$ in Lemma~\ref{lemma: reflection_conv} extends to $\mathcal{P}_3$. Despite the introduction of two variable sets in $\mathcal{P}_3$, $\mathbf{p}_a$ and $\mathbf{p}_s$, convexity remains intact due to their linear and independent impact on the objective function and constraints. This linear contribution preserves $\mathcal{P}_3$'s convexity, akin to the structure established in Lemma~\ref{lemma: reflection_conv}.
\end{proof}
\vspace{-2mm}

As the power control problem $\mathcal{P}_{3}$ is convex after FP transformation, the optimal solution can be found by utilizing optimization tools such as \texttt{MATLAB} coupled with \texttt{CVX}, a package for specifying and solving convex programs implemented with the Interior Point Method for guaranteeing convergence to the global optimum. The algorithmic implementation of this method is delineated in Algorithm~\ref{Algo:power}.
\begin{algorithm}[!t]
\setstretch{1.38}
	{\small\caption{Power control optimization}\label{Algo:power}
		\begin{algorithmic}[1]
			\Require 
			\parbox[t]{\dimexpr\linewidth-\algorithmicindent-\algorithmicindent\relax}{$\alpha, \beta [n], s_{tg}[n], x_g[n], d_{te}[n], D[n], \forall n \in \mathcal{N}$, $L$, $x_t$, $\sigma_r^2$, $\sigma_e^2$, $K_{tg}$, $P$, $i$, and tolerance $\xi$.\strut}
			\Ensure 
			\parbox[t]{\dimexpr\linewidth-\algorithmicindent-\algorithmicindent\relax} {Optimal AN and CW power, $p_a [n]^*, p_s [n]^* \forall n \in \mathcal{N}$.\strut}
			\State Set $\mathrm{it}=1,\, p_a [n]^{\rm(it)} = p_s [n]^{\rm(it)} = \frac{P}{2N}, \forall n \in \mathcal{N}$.
			\State Calculate $\Lambda^{(\mathrm{1})} = \sum_{n = 1}^N R [n]$ by substituting $P_a [n]^{\rm(it)}$ and $P_s [n]^{\rm(it)}, \forall n \in \mathcal{N}$. 
			\Do\Comment{Iteration}
   \vspace{-1mm}
			\State Set $\mathrm{it}=\mathrm{it}+1$.
   
            \State\parbox[t]{\dimexpr\linewidth-\algorithmicindent-\algorithmicindent\relax}{Set $y_1\hspace{-0.7mm}[\hspace{-0.3mm}n\hspace{-0.3mm}]$ and $y_2 \hspace{-0.4mm}[\hspace{-0.3mm}n\hspace{-0.3mm}]$ by substituting $\beta \hspace{-0.3mm}[\hspace{-0.3mm}n\hspace{-0.3mm}]\hspace{-0.4mm}^{(\hspace{-0.3mm}\mathrm{it}\hspace{-0.3mm})}$ into~\eqref{y1}-\eqref{y2}. \strut}
            
			\State\parbox[t]{\dimexpr\linewidth-\algorithmicindent-\algorithmicindent\relax} {Solve $\mathcal{P}_{3}$ using \texttt{CVX} and return $p_a\hspace{-0.3mm}[\hspace{-0.3mm}n\hspace{-0.3mm}] \hspace{-0.3mm}^{\rm(\hspace{-0.3mm}it\hspace{-0.3mm})}\hspace{-0.8mm}, p_s\hspace{-0.3mm}[\hspace{-0.3mm}n\hspace{-0.3mm}] \hspace{-0.3mm}^{\rm(\hspace{-0.3mm}it\hspace{-0.3mm})}\hspace{-0.8mm}.$\strut}
			\State 
			\parbox[t]{\dimexpr\linewidth-\algorithmicindent-\algorithmicindent\relax}{Calculate $\Lambda^{(\mathrm{it})}\hspace{-0.5mm} = \hspace{-0.5mm}\sum_{n = 1}^N \hspace{-0.5mm}R [n]$ by substituting $p_a [n]^{\rm(it)}$ and $p_s [n]^{\rm(it)}, \forall n \in \mathcal{N}$\strut} 
			\doWhile{$\left(\Lambda^{(\mathrm{it})}-\Lambda^{(\mathrm{it}-1)}\right)\le\xi$}\Comment{Termination}
			\State \textbf{Return}  ${p_a [n]^{*}}={p_a [n]^{\rm(it)}}$, and ${p_s [n]^{*}}={p_s [n]^{\rm(it)}}$. 
				
		\end{algorithmic}
	}
\end{algorithm}

\subsection{TV's trajectory optimization}
With the optimal reflection coefficients and power allocation established in Sections~\ref{sec:bate_opt} and~\ref{sec:Power_opt}, here we present the design of the TV's linear trajectory. It may be noted that the secrecy rate for each slot only depends on its location for that particular slot. Thus, we can design the trajectory slot by slot.
Each $x_g [n]$ depends on $x_g [n-1]$ with the velocity and acceleration constraints.
The trajectory optimization problem is then formulated as follows:
\begin{align*}
&\mathcal{P}_4: \hspace{-0.5mm}\underset{x_g[n]}{\text{maximize}}\hspace{0.5mm} R [n], \hspace{0.5mm} 
\text{s.t.}: ({\rm C4}) - ({\rm C7}).
\end{align*}
We simplify our notation by letting $A[n] = \frac{1+\gamma_t[n]}{1+\gamma_e[n]}, n \in \mathcal{N}$. Taking the derivative of $A[n]$ and setting it to zero, we can obtain an $11$th order polynomial equation and find all the solutions. Thus, it is categorized as a particular class of optimization problems characterized by a finite set of optimal points.  
This finite nature simplifies the search for the optimal trajectory by confining it to candidate solutions.

The $11$ candidate points are subject to the constraints $({\rm C4}) - ({\rm C7})$. Here, we introduce Theorem~\ref{Theorem:points} to establish the conditions for global optimality. Specifically, the theorem examines the intersection of potential optimal points and the active constraint boundaries within the solution space of $\mathcal{P}_{4}$ to find feasible solutions.
\begin{theorem}\label{Theorem:points}
    The global optimal trajectory $x_g^* [n], \forall n \in \mathcal{N}$, maximizing $R [n](x_g [n])$ over the set of all optimal points of $\mathcal{P}_4$, is given by~\eqref{solution}.
\end{theorem}
\vspace{-3mm}
\begin{proof}
    Given the boundary conditions $({\rm C6})$ and $({\rm C7})$, the extremities $x_g^{\text{min}} [n]$ and $x_g^{\text{max}} [n]$ are established as critical corner points. Coupled with the identification of eleven potential critical or gradient points, this comprehensive set encompasses all optimal solutions as delineated in~\eqref{solution}.
\end{proof}
\vspace{-5mm}
\begin{equation} \label{solution}
\begin{aligned}
    x_g^* [n] &= \underset{x_g[n]}{\text{argmax}} R[n]\left(\left[\{x_g^{\text{min}}[n],x_g^{\text{max}}[n]\}: \text{Corner}\right], \right.\\ 
    &\left. \left[\{x_g^{r1}[n], x_g^{r2} [n], ..., x_g^{r11} [n]\}: \text{Gradient}\right]\right)
\end{aligned}
\end{equation}


\section{Numerical Performance Evaluation}\label{sec:results}
In this section, we conduct numerical simulations to verify the analysis and provide key design insights, including the performance gain achieved over the existing benchmark. 
\subsection{Default Simulation Parameters}
Default system parameters are as follows unless stated otherwise: lane separation is $3.5\,\text{m}$, $t = 0.05\,\text{s}$, $\sigma_e^2 = \sigma_r^2 = -80\,\text{dBm}$, $\alpha = 0.5$, $P = 20\,\text{W}$, $i = 2$ as indicated in~\cite{pathloss}, and a fixed distance-independent path loss constant at $0.2$. The EV travels at $v_c = 30\,\text{m/s}$, covering distance $L = Nv_c$ over $N$ time slots, and $t = \left(\frac{L}{2}, 8, 3\right)$. TV's velocity varies between $[17, 40]$ m/s, with acceleration bounds $\left[-5, 5\right]\,\text{m/s}^{2}$. The goal is to evaluate the aggregated secrecy rate over successive $N$ time slots to simulate the vehicle's dynamics.

\begin{figure*}[!t]	
	\begin{minipage}{.28\textwidth}	
		\centering\includegraphics[width=1.75in]{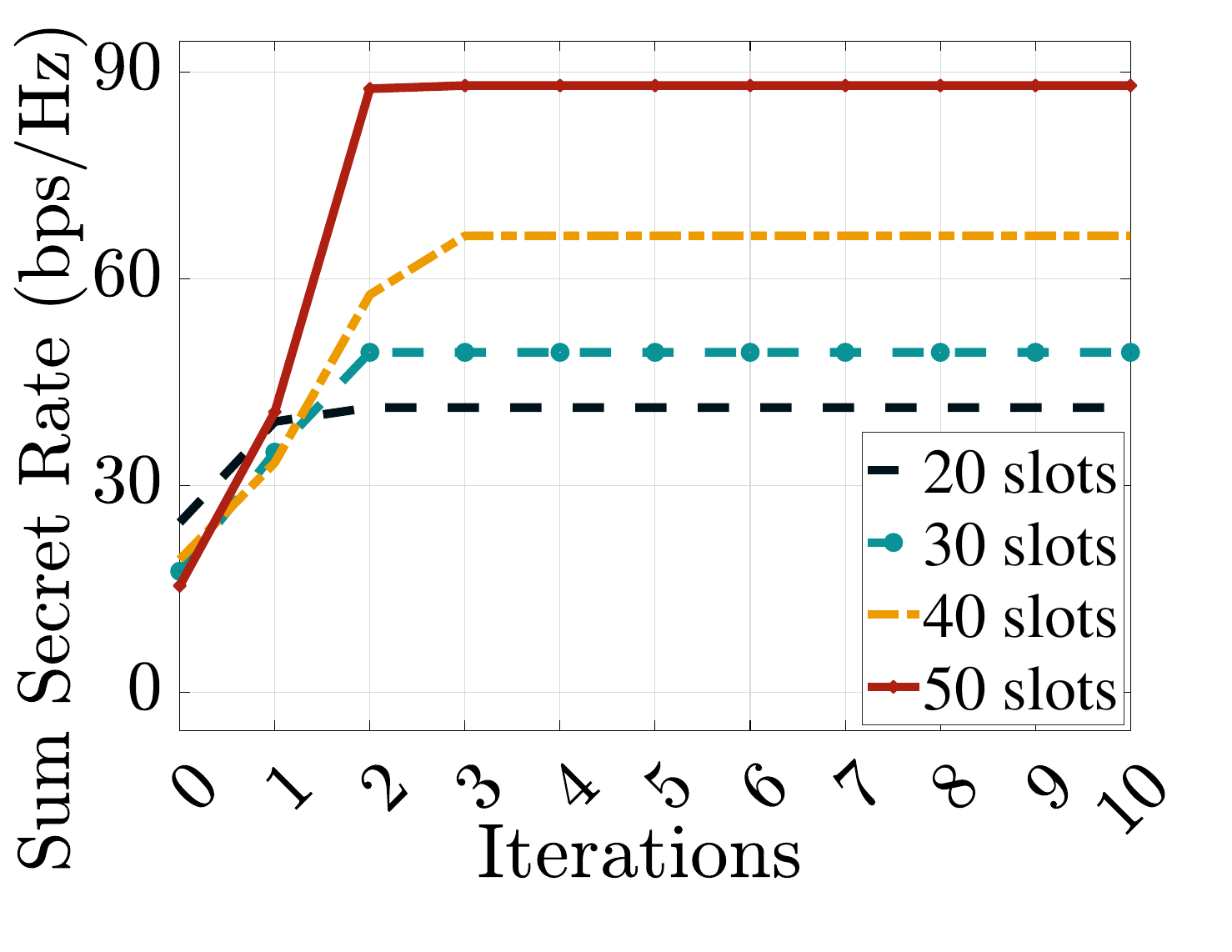}  
    \caption{Comparison of convergence for the proposed optimization framework with different numbers of time slots under the same power budget and the same coherent time}
    \label{fig:iteration} 
	\end{minipage}\quad 
	\begin{minipage}{.44\textwidth}	
		\subfigure[Coincident TV-EV Separation]{
		\includegraphics[width=1.45in]{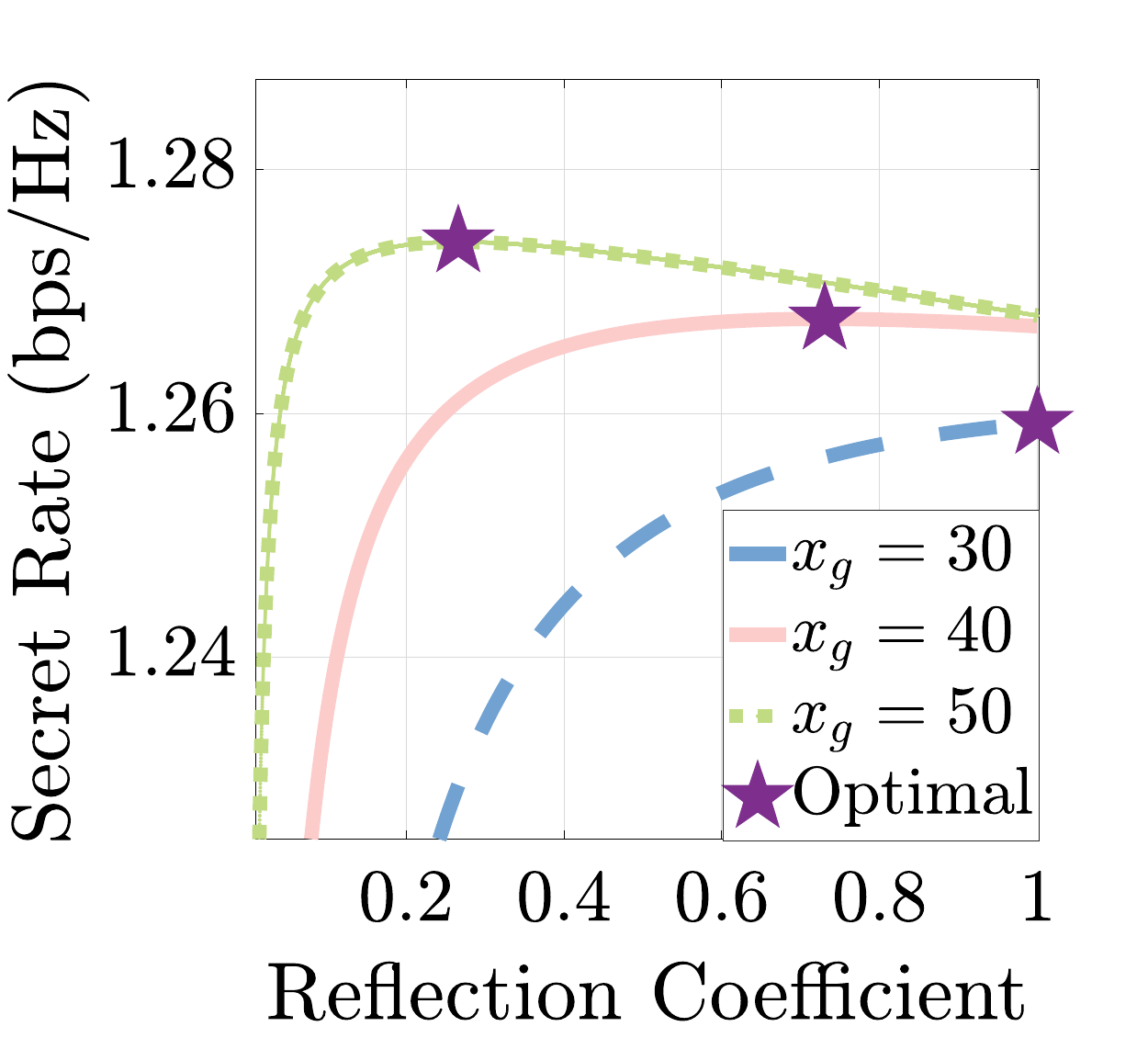}
	}\hspace{-2mm}
	\subfigure[30m TV-EV Separation]{
		\includegraphics[width=1.45in]{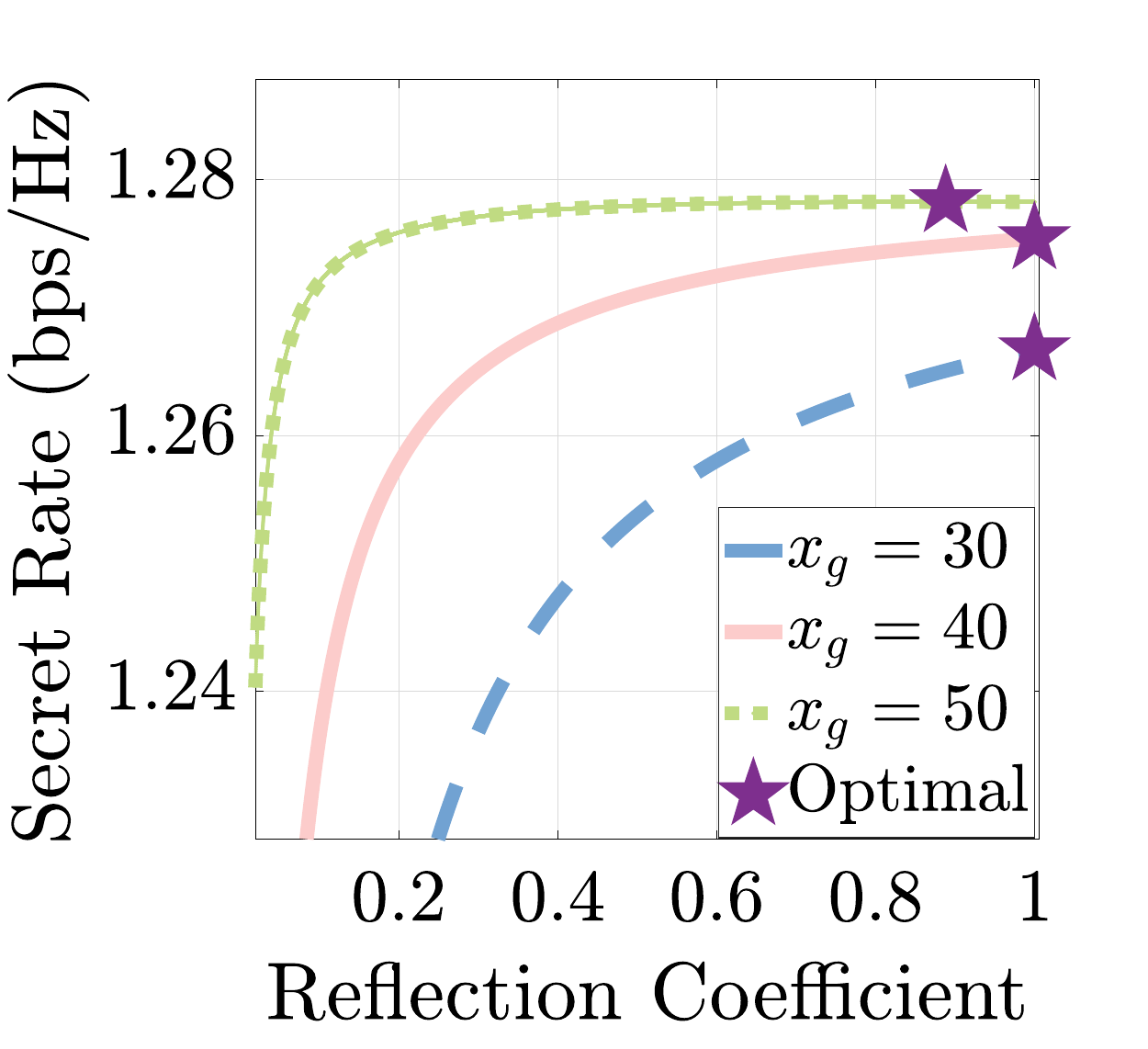}
	}
	\caption{Validating the global optimality of the proposed reflection coefficient optimization algorithm}
	\label{Fig: reflection}	
	\end{minipage}\;
	\begin{minipage}{.2\textwidth}	
		\centering\includegraphics[width=1.65in]{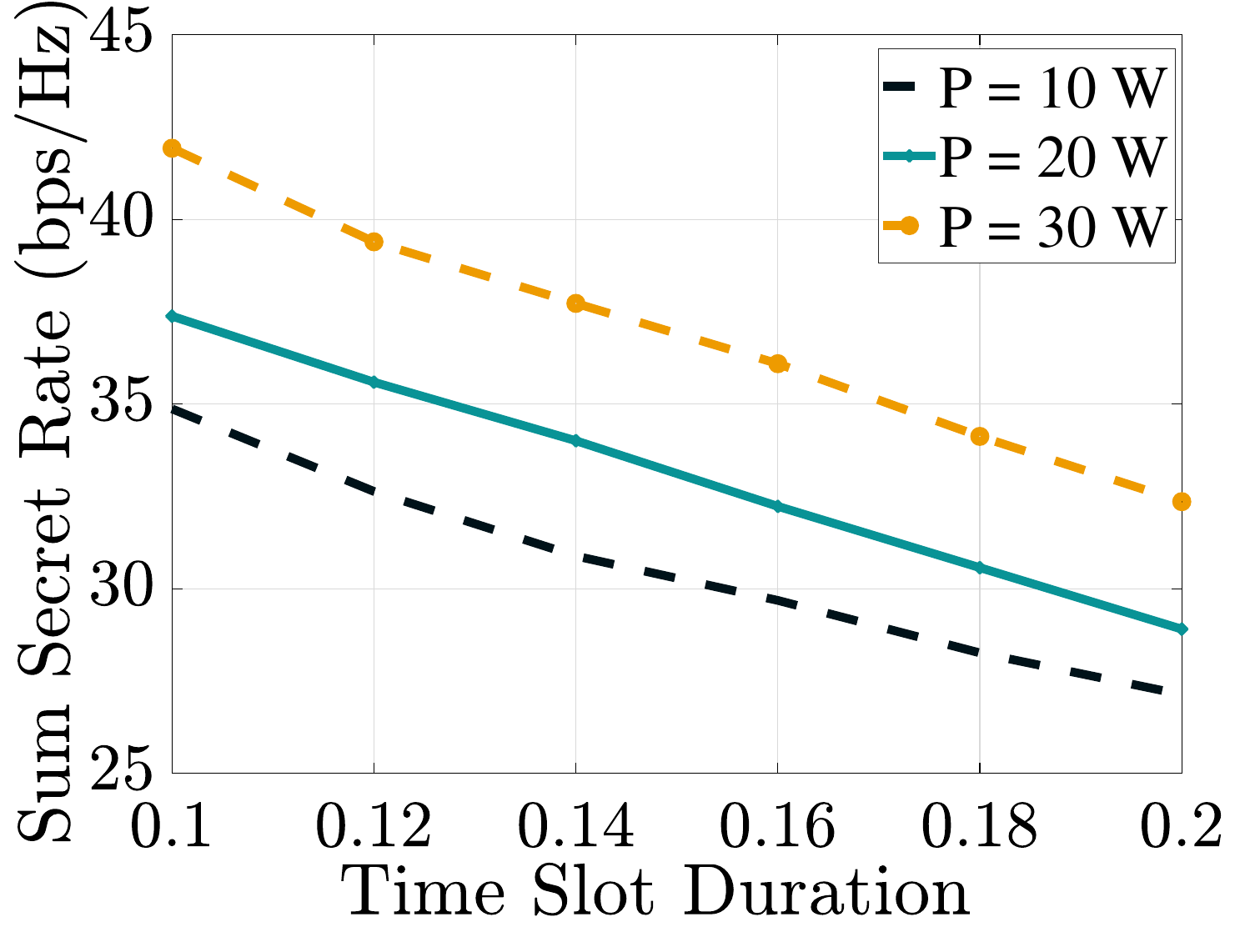}
\caption{Sum secrecy rate comparison across varying time slot durations for different power budgets with \( N = 20 \) nodes.}
\label{fig: timeSlots}
	\end{minipage}\vspace{-4mm} 
\end{figure*}  

\subsection{Convergence discussion and optimal validation}
We examine the efficacy of our proposed joint optimization algorithm compared to a default baseline across various time slots. Figure~\ref{fig:iteration}  illustrates the evolution of the sum secrecy rate across $10$ iterations for the proposed joint optimization algorithm and demonstrates this has a rapid convergence benefit. Specifically, systems with fewer time slots tend to converge more rapidly; those with $N = 20$ and $N = 30$ require only $2$ iterations to converge, while others necessitate $3$ iterations. Additionally, it can be observed that the starting sum secrecy rate with $20$ time slots outperforms the others, which can be attributed to the average proximity between the TV and infrastructure being minimized when $N = 20$. In scenarios devoid of optimization strategies, the received power at the TV dominates the secrecy rate. 
Moreover, the communication security is significantly enhanced after implementing the proposed algorithm across all configurations. The improvements are quantified at $0.68$, $1.81$, $2.44$, and $4.71$ times for $N = 20$, $N = 30$, $N = 40$, and $N =50$ from $24.60, 14.54, 19.24, \text{and} \ 15.43$ bps/Hz respectively. Increasing the number of time slots provides more opportunities for improvement despite the constant power budget.

In Fig.~\ref{Fig: reflection}, we validate the proposed reflection coefficient optimization algorithm by examining the secrecy rate variation against increasing reflection coefficients. The scenario maintains a constant configuration, with the infrastructure at \( t = (50, 8, 3) \). In Fig.~\ref{Fig: reflection} (a), the eavesdropper aligns with the TV's position, whereas it is 30 m away in Fig.~\ref{Fig: reflection} (b).
The temporal specificity of our analysis does not preclude the generality of our findings, as the observed trends are consistent across different time slots. 
Both subplots in Fig.~\ref{Fig: reflection} reveal an unimodal relationship between the secrecy rate and the reflection coefficient, where the peak corroborates the global optimality of our algorithm. 
Notably, the optimal reflection coefficient decreases with the closer distance between TV and infrastructure, improving the secrecy rate. It confirms the observation from Fig.~\ref{fig:iteration} that the distance between infrastructure and TV is significant for the secrecy rate. 
Furthermore, a comparative analysis of the subplots in Fig.~\ref{Fig: reflection} elucidates that the tag is inclined to augment reflection when the eavesdropper is farther from the TV. This observation provides evidence of the system's capability to mitigate eavesdropping threats.

\subsection{Optimal Design Insights and performance camparasion}
In this subsection, we provide insights into this proposed joint algorithm and conduct a comparative analysis of the contribution of each optimized variable to the overall system efficacy. 
We start by comparing the sum secrecy rates over $N = 20$ time slots for various power budgets with $t$ varying from $0.1$s to $0.2$s, as shown in Fig.~\ref{fig: timeSlots}.
The simulation results exhibit a clear trend where an increase in the power budget leads to a rise in the sum secrecy rate. For power budgets of $P = 10\,\text{W}$, $P = 20\,\text{W}$, and $P = 30\,\text{W}$, the corresponding average sum secrecy rates are $30.58\,\text{bps/Hz}$, $33.11\,\text{bps/Hz}$, and $36.94\,\text{bps/Hz}$. The improvement is because of the greater flexibility in power allocation afforded by a higher power budget.
Conversely, a longer time slot ($t = 0.2$ s) results in a $29.17\%$ reduction in the average sum secrecy rate compared to $t = 0.1$ s for the same power budgets in Fig.~\ref{fig: timeSlots}. This decline is ascribed to the more considerable distances the vehicle covers in extended time slots, impacting trajectory design accuracy and, consequently, the secrecy rate. Additionally, an extended $t$ increases the average distance between infrastructure and TV, further decreasing the secrecy rate.

\begin{figure}[h] 
	\centering
 \vspace{-3pt}
	\subfigure[Secrecy Rate vs. Distance]{
		\includegraphics[width=1.253in]{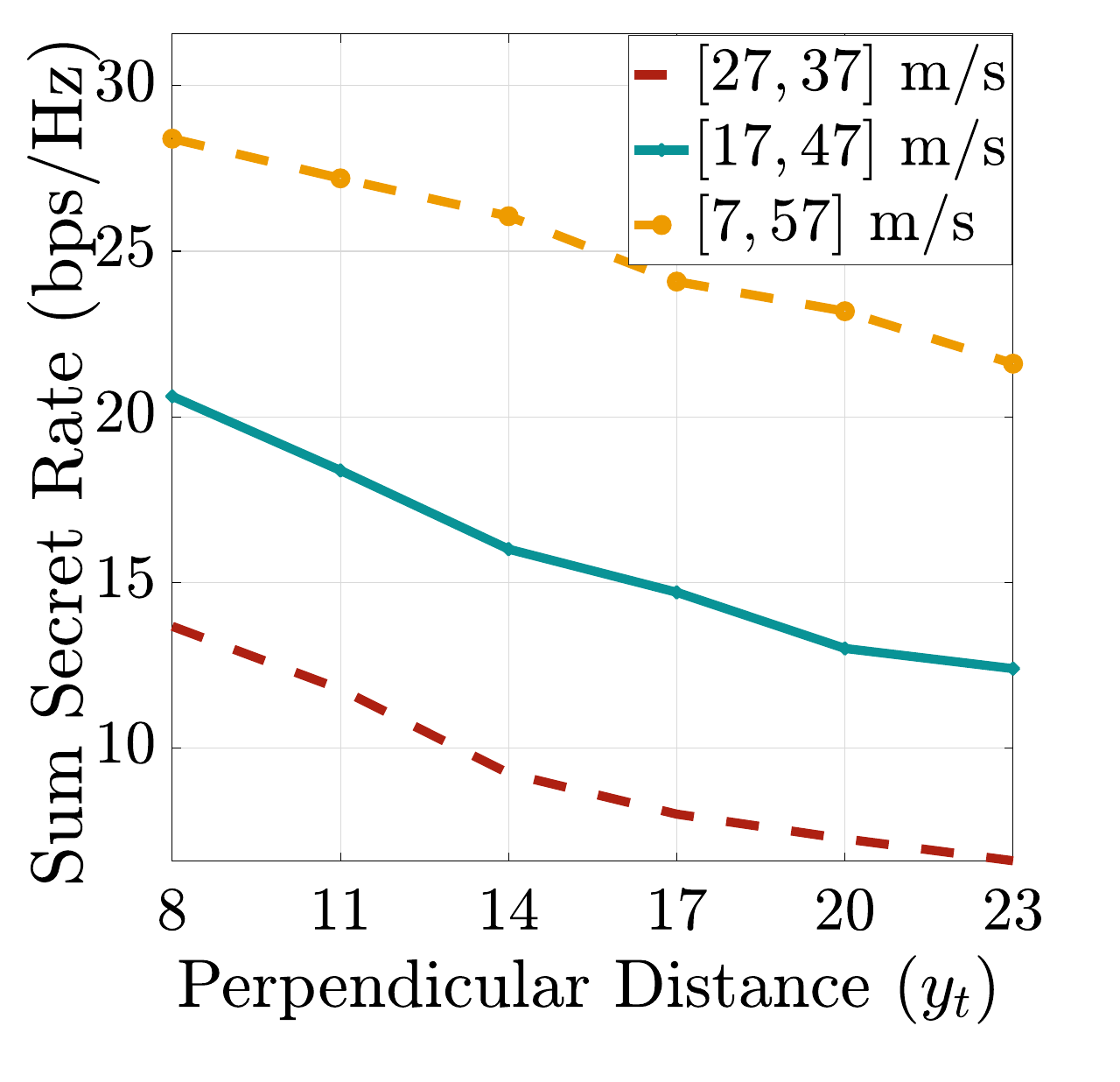}
	}\hspace{-3mm}
	\subfigure[Variables Compression]{
		\includegraphics[width=2.08in]{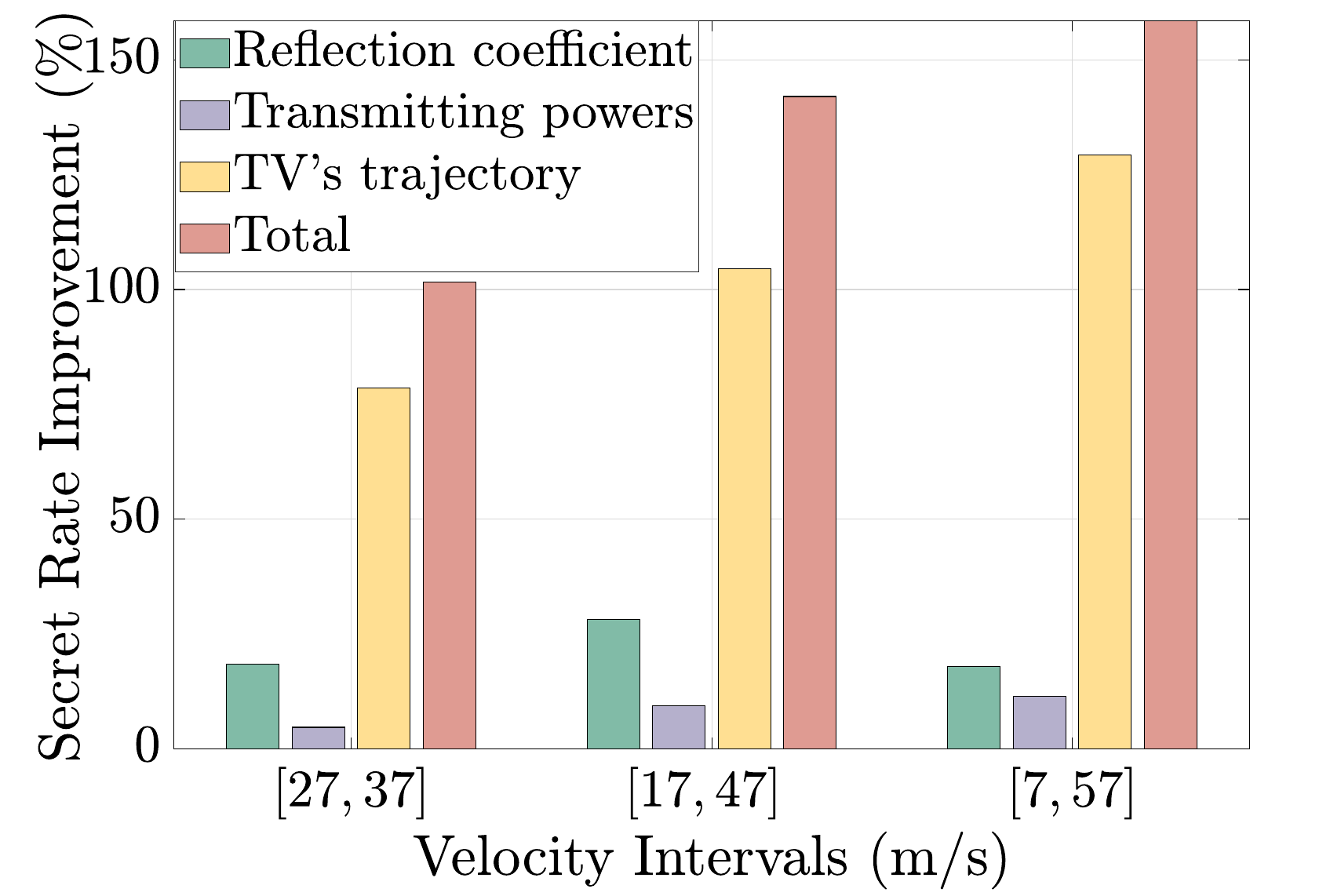}
	}
	\caption{Comparison of the sum secrecy rate across varying perpendicular distances to the road under different velocity constraints, with \( N = 20 \) nodes, alongside the analysis of objective improvement for various variables}
	\label{Fig:comparasion}	
\end{figure}
In our comparative analysis, we evaluate the impact of each optimized variable on system efficacy, focusing on the infrastructure's positioning relative to the roadside. Fig.~\ref{Fig:comparasion} (a) illustrates the relationship between varying perpendicular distances from the road and the sum secrecy rate under different vehicular speed constraints. Fig.~\ref{Fig:comparasion} (b) further analyzes the individual contributions of each optimized variable to the sum secrecy rate, starting from the initial distance presented in Fig.~\ref{Fig:comparasion} (a).
It can be observed from subplot (a) that a decrease in the objective function with increased distance is primarily due to heightened path loss. Also, broader velocity intervals lead to a higher sum secrecy rate as it provides a more flexible trajectory design. It is confirmed by Fig.~\ref{Fig:comparasion} (b), in which the expanded velocity boundary yields trajectory optimization improvements of $4.64\%$, $9.40\%$, and $11.42\%$, attributed to the greater latitude in trajectory design.
It is essential to highlight that power allocation significantly influences system performance in this joint optimization context, leading to an average enhancement of $104.16\%$. Subsequently, reflection coefficient optimization and trajectory design contribute notably, with moderate improvements of $21.46\%$ and $8.49\%$, respectively. These findings provide valuable insights for optimization strategies in related future research endeavours.

\section{Concluding Remarks}\label{sec:Conclusion}
This study investigates a novel green V2I backscattering system to maximize PLS with a low-complexity approach. We adopt the AO method to decompose the problem, ensuring the convexity of sub-problems through parallel computing, FP transformation, and feasible analysis. It facilitated achieving global optima for each subproblem and a near-global optimal for the principle problem. Comprehensive numerical analyses substantiated our theoretical models, provided valuable insights for future study, and demonstrated a significant elevation in the sum secrecy rate. Specifically, for \( N = 50 \), the proposed scheme achieved a $4.7$ times improvement in performance relative to conventional benchmarks. 

\vspace{4pt}
\section*{Acknowledgment} 
This research was supported by Australian Research Council Discovery Early Career Researcher Award - DE230101391.

\vspace{4pt}
\end{document}